\documentclass[webpdf,contemporary,large]{oup-authoring-template}%

\usepackage{bbm}
\usepackage{amsmath}

\graphicspath{{Fig/}}

\algrenewcommand\algorithmicrequire{\textbf{Input:}}
\algrenewcommand\algorithmicensure{\textbf{Output:}}

\theoremstyle{thmstyleone}%
\theoremstyle{thmstyletwo}%
\theoremstyle{thmstylethree}%

\begin{document}

\journaltitle{Journal Title Here}
\DOI{DOI HERE}
\copyrightyear{2022}
\pubyear{2019}
\access{Advance Access Publication Date: Day Month Year}
\appnotes{Paper}

\firstpage{1}


\title[Integrative Analysis of Multiple Data Streams with DGP-SI]{Integrative Analysis and Imputation of Multiple Data Streams via Deep Gaussian Processes}

\author[1,$\ast$]{Ali A. Septiandri\ORCID{0000-0003-2044-5106}}
\author[2]{Deyu Ming\ORCID{0000-0003-2369-1927}}
\author[3]{F. A. Diaz De la O\ORCID{0000-0001-6277-7788}}
\author[1]{Takoua Jendoubi\ORCID{0000-0001-7846-9763}}
\author[4]{Samiran Ray\ORCID{0000-0002-3738-4672}}

\authormark{Septiandri et al.}

\address[1]{\orgdiv{Department of Statistical science}, \orgname{University College London}, \orgaddress{\street{London} \postcode{WC1E 7HB}, \country{UK}}}
\address[2]{\orgdiv{School of Management}, \orgname{University College London}, \orgaddress{\street{London} \postcode{WC1E 6BT}, \country{UK}}}
\address[3]{\orgdiv{Clinical Operational Research Unit}, \orgname{University College London}, \orgaddress{\street{London} \postcode{WC1H 0BT}, \country{UK}}}
\address[4]{\orgdiv{Paediatric Intensive Care Unit}, \orgname{Great Ormond Street Hospital For Children NHS Foundation Trust}, \orgaddress{\street{London} \postcode{WC1N 3BH}, \country{UK}}}

\corresp[$\ast$]{Corresponding author. \href{email:ali.septiandri.21@ucl.ac.uk}{ali.septiandri.21@ucl.ac.uk}}

\received{Date}{0}{Year}
\revised{Date}{0}{Year}
\accepted{Date}{0}{Year}



\abstract{\textbf{Motivation:} Healthcare data, particularly in critical care settings, presents three key challenges for analysis. First, physiological measurements come from different sources but are inherently related. Yet, traditional methods often treat each measurement type independently, losing valuable information about their relationships. Second, clinical measurements are collected at irregular intervals, and these sampling times can carry clinical meaning. Finally, the prevalence of missing values. Whilst several imputation methods exist to tackle this common problem, they often fail to address the temporal nature of the data or provide estimates of uncertainty in their predictions. \\
\textbf{Results:} We propose using deep Gaussian process emulation with stochastic imputation, a methodology initially conceived to deal with computationally expensive models and uncertainty quantification, to solve the problem of handling missing values that naturally occur in critical care data. This method leverages longitudinal and cross-sectional information and provides uncertainty estimation for the imputed values. Our evaluation of a clinical dataset shows that the proposed method performs better than conventional methods, such as multiple imputations with chained equations (MICE), last-known value imputation, and individually fitted Gaussian Processes (GPs). \\
\textbf{Availability and implementation:} The source code of the experiments is freely available at: \url{https://github.com/aliakbars/dgpsi-picu}. \\
\textbf{Contact:} \href{ali.septiandri.21@ucl.ac.uk}{ali.septiandri.21@ucl.ac.uk}
}


\maketitle

\section{Introduction}
\label{sec:intro}

One of the main challenges in analysing healthcare data is that they are usually collected from multiple measurement streams. A patient's medical record may include data from different sources, such as medical histories, laboratory tests, and imaging studies~\cite{johnson2016mimic}. Integrating and analysing the data can thus be challenging, as the various data sources may use different formats, units, and scales. For example, a patient's CO$_2$ level may be measured breath by breath directly from a ventilator circuit, but their albumin levels are measured daily from blood sample tests. Thus, aligning these multiple streams will result in missing values. Simply removing missing values and performing a complete case analysis would not suffice, since useful observations might be lost~\cite{vesin2013missing}.

Another challenge of working with healthcare data is that it is often irregularly and informatively sampled~\cite{Groenwold2020Dec}. The irregularity refers to the data collection at different times and intervals due to the sample sources. Additionally, there is the problem of informative sampling, where one might observe an extended period of intervals because the patient is getting better, making the clinicians reduce the frequency of monitoring~\cite{che2018recurrent}. These circumstances make it challenging to assess a patient's health and make informed decisions accurately. As the sampling frequency is, by nature, informative, it will be more difficult to detect subtle changes in unobserved variables retrospectively.

Current practices for handling missing values in healthcare data often prioritise simplicity over complexity. A common approach is using the last-known value imputation, also known as the last-observation-carried-forward (LOCF) method, which fills in missing values by extending the most recent available measurement for a specified time period~\cite{siddiqui1998locf}. While this method is straightforward, it overlooks the correlation between covariates. On the other hand, multiple imputation by chained equations (MICE), another commonly widely used method, uses cross-sectional information between covariates but treats each observation independently~\cite{rubin1987mice,tsvetanova2021inconsistent}.

More recently, deep neural networks have become increasingly popular for handling missing values in healthcare data~\cite{lipton16rnn,che2018recurrent,cao2018brits,yildiz2022multivariate}. However, these methods have a limitation: they typically do not provide estimates of uncertainty in their predictions. This can be a problem in healthcare data analysis, where medical observations and interpretations inherently contain uncertainty, which may come from measurement error, inherent noise in the signal, or the use of surrogate markers. When this uncertainty in the input data is not accounted for, it can lead to unreliable model predictions~\cite{cabitza2017unintended}.

This study aims to tackle the problem of handling missing values in multivariate time series data by leveraging both longitudinal and cross-sectional information. We use a deep Gaussian process (GP) model with stochastic imputation~\cite{ming2021linked,ming2023deep} where time is the input to predict the target variable through covariates in the model's latent layer. By fitting the model jointly, the available information can be used as leverage to impute missing values stochastically. Moreover, this method comes with uncertainty estimation. This approach is evaluated against a baseline method that fits individual GPs to covariates using complete case analysis for model training and subsequent imputation.

A deep Gaussian process model is a hierarchical structure of GP nodes organised in layers to represent latent variables~\cite{damianou2013deep}. Each node receives input from the previous layer and produces output that serves as input for the next layer. The observed data points appear at the final layer of this hierarchy. While single-layer GP models are limited by the kernel function used, which can be highly parameterised to learn complex data patterns, a deep GP model learns them non-parametrically via the hierarchy, thus having fewer hyperparameters to optimise~\cite{salimbeni2017doubly}. Due to their ability to provide uncertainty estimates, deep GP models have applications in real-life domains, including aero-propulsion system simulation~\cite{biggio2021uncertainty}, crop yield prediction from remote sensing data~\cite{you2017deep}, and uncertainty estimation in electronic health records~\cite{li2021deep}.

The remainder of the paper is structured as follows. The clinical problem that motivated this research is explained in Section~\ref{sec:related}. GPs and deep GPs are reviewed in Section~\ref{sec:methodology},  where the methodological approach and alternative techniques for handling missing values in multivariate time series data are described. The proposed deep GP using stochastic imputation is validated by applying it to a clinical case study in Section~\ref{sec:results}. Finally, findings and future directions are summarised in Section~\ref{sec:conclusion}.
\section{Motivation: Clinical Problem}
\label{sec:related}

Missing values are a common challenge in healthcare datasets, arising from sources such as incomplete patient forms, survey non-responses~\cite{penny2012approaches}, and technical glitches during data collection~\cite{zhang2019defects}. These missing values manifest within the data and affect study outcomes and statistical validity. Therefore, choosing appropriate methodological approaches to tackle this problem is crucial.

In critical care medicine, missing values are also a result of irregular and informative sampling~\cite{Groenwold2020Dec}. To provide some context, clinicians in critical care monitor deviations from the expected arterial acidity (pH) range to gain insights into respiratory function, electrolyte balance, and the underlying diseases of the patients~\cite{sirker2002acid}. When pH deviates from normal ranges, either through acidosis (pH $< 7.3$) or alkalosis (pH $> 7.5$), it could disrupt vital biochemical processes and overall equilibrium, with studies showing that blood pH levels are associated with the mortality rate~\cite{rodriguezvillar2021acidemia,jung2011acidemia} and neurological recovery in cases of cardiopulmonary resuscitation~\cite{shin2017ph}.

One way to monitor and model the pH level is the Stewart-Fencl approach~\cite{stewart1978variables,stewart1983modern,fencl1993stewart}. This approach identifies three independent variables that determine pH. The first variable is carbon dioxide (CO$_2$), a major source of acid in the body that can be continuously monitored using modern bedside equipment. The second and third variables are strong ion differences (e.g. Na$^+$, K$^+$, Cl$^-$) and weak acids (e.g. albumin, lactate, urea, phosphate), respectively. These components require blood tests for measurement, which are performed less frequently due to their invasive nature~\cite{barie2004phlebotomy}.

While CO$_2$ is also measured through capnography as a surrogate, known as end-tidal CO$_2$ (ETCO$_2$), the main interest is in blood CO$_2$ levels because they directly influence pH. In various respiratory conditions (such as asthma and COPD), the CO$_2$ in the blood does not equilibrate with CO$_2$ in the lungs. This creates a measurable gap between blood and alveolar CO$_2$ levels, which can also be informative~\cite{anderson2000co2,lee2024pitfalls}.

\begin{figure}[htbp]
    \centering
    \includegraphics[width=\linewidth]{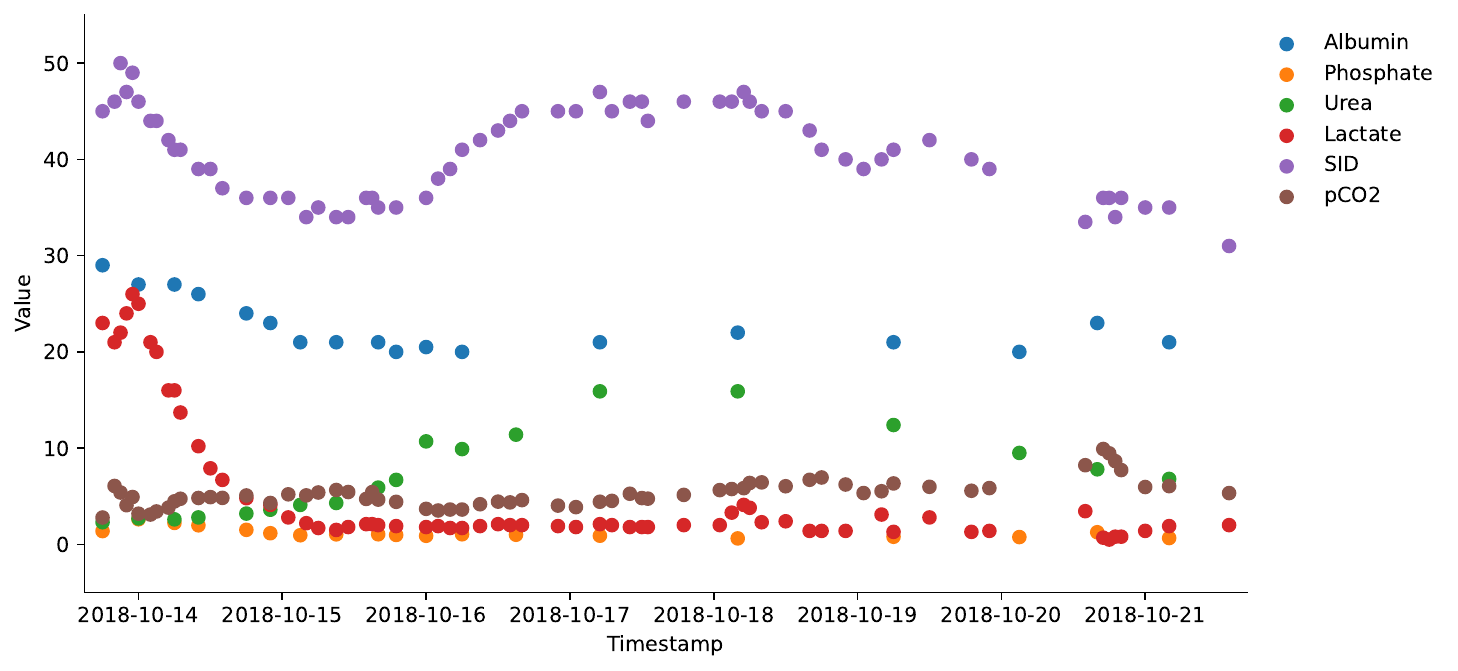}
    \caption{Irregular sampling of six measurements of a sample patient. The pCO$_2$, strong ion differences (SID), and lactate are measured from bedside monitoring, while albumin, phosphate, and urea levels are obtained from blood tests.}
    \label{fig:sampling}
\end{figure}

This disparity in measurement frequency creates a pattern of irregular sampling, where data collection occurs at inconsistent intervals. The resulting gaps in data collection lead to missing values, particularly for parameters requiring blood tests, as illustrated in Fig.~\ref{fig:sampling}. Consequently, the irregular nature of these measurements complicates the application of standard statistical techniques to analyse and interpret the data.

On the other hand, informative sampling occurs when data point selection is influenced by factors related to clinical hypotheses. For instance, clinicians might refrain from collecting data when a patient's health improves and vice versa~\cite{che2018recurrent}. In this context, both irregular and informative sampling scenarios fall under the Missing Not at Random (MNAR) category, as the missingness is not random but associated with unobserved data or specific conditions and the missingness carries information~\cite{little2019statistical}.

This study addresses the challenge of missing values in critical care data, which can impact patient outcome predictions~\cite{vesin2013missing}, for example, in predicting in-hospital and 30-day mortality~\cite{sharafoddini2019missing}. Rather than relying on complete-case analysis, which risks losing valuable information, this work proposes using deep GPs for data imputation. This approach leverages both cross-sectional and longitudinal information from patient records.

Following the Stewart-Fencl approach, the analysis is structured with arterial pH as the dependent variable.  Relevant covariates are constructed to identify factors causing pH deviations. The proposed method not only imputes missing values in the covariates but also quantifies the uncertainty associated with these imputations, providing a more comprehensive understanding of the data's reliability.

While this study focuses on its application in critical care medicine because of its clinical importance, the proposed method applies to different scenarios with similar conditions. For example, missing values are also found in human activity recognition from multiple sensor streams~\cite{jain2022collossl}, sleep disorder diagnoses using electroencephalogram (EEG)~\cite{lee2021contextual}, and hepatocellular carcinoma~\cite{han2021hepatocellular}.
\section{Methodology}
\label{sec:methodology}

\subsection{Gaussian Processes}
\label{subsec:gp}

Let $\mathbf{X} \in \mathbb{R}^{N \times D}$ represent a $D$-dimensional input with $N$ observed data points and $\mathbf{Y} \in \mathbb{R}^{N}$ be the corresponding outputs. Then, the GP model assumes that $\mathbf{Y}$ follows a multivariate Gaussian distribution $\mathbf{Y} \sim \mathcal{N}(\boldsymbol{\mu}, \sigma^2 \mathbf{R}(\mathbf{X}))$, where $\boldsymbol{\mu} \in \mathbb{R}^{N}$ is the mean vector, $\sigma^2$ is the scale parameter, and $\mathbf{R}(\mathbf{X}) \in \mathbb{R}^{N \times N}$ is the correlation matrix. Cell $ij$ in the matrix $\mathbf{R}(\mathbf{X})$ is specified by $k(\mathbf{X}_{i*}, \mathbf{X}_{j*}) + \eta \mathbbm{1}_{\{\mathbf{X}_{i*} = \mathbf{X}_{j*}\}}$, where $k(\cdot, \cdot)$ is a given kernel function with $\eta$ being the nugget term and $\mathbbm{1}_{\{\cdot\}}$ being the indicator function. In this study, we consider Gaussian processes with zero means, i.e. $\boldsymbol{\mu} = \mathbf{0}$ and kernel functions with the multiplicative form: $k(\mathbf{X}_{i*}, \mathbf{X}_{j*}) = \prod_{d=1}^D k_d(X_{id}, X_{jd})$ where $k_d(X_{id}, X_{jd}) = k_d(|X_{id} - X_{jd}|)$ is a one-dimensional stationary and isotropic kernel function, e.g., squared exponential kernel function~\cite{rasmussen2006gaussian}, for the $d$-th input dimension.

The hyperparameters $\sigma^2$, $\eta$, and those contained in $k(\cdot, \cdot)$ are typically estimated using maximum likelihood or maximum a posteriori~\cite{rasmussen2006gaussian}. Given estimated GP hyperparameters, the realisations of input $\mathbf{x} = (\mathbf{x}_{1*}^T, ..., \mathbf{x}_{N*}^T)^T$, and output $\mathbf{y} = (y_1, ..., y_N)^T$, then the posterior predictive distribution of output $Y_0$ at a new input position $\mathbf{x_0} \in \mathbb{R}^{1 \times D}$ follows a Gaussian distribution with mean $\mu_0$ and variance $\sigma_0^2$ given by:
\begin{align}
    \mu_0 &= \mathbf{r}(\mathbf{x}_0)^T \mathbf{R}(\mathbf{x})^{-1}\mathbf{y} \\
    \sigma_0^2 &= \sigma^2 (1 + \eta - \mathbf{r}(\mathbf{x}_0)^T \mathbf{R}(\mathbf{x})^{-1} \mathbf{r}(\mathbf{x}_0))
\end{align}
where $\mathbf{r}(\mathbf{x}_0) = [k(\mathbf{x}_0, \mathbf{x}_{1*}), ..., k(\mathbf{x}_0, \mathbf{x}_{N*})]^T$.

A GP model can be used as a smoothing function for irregularly sampled signals through the predicted mean function of a time series. Thus, GPs have been used to model electronic health records~\cite{lasko2013smoothing}, wearable sensor data for e-health~\cite{clifton2012gaussian}, gene expression data~\cite{gao2008gp,kirk2009gp,liu2010gene}, and quantitative traits~\cite{vanhatalo2019gp,arjas2020snp}.

\subsection{Linked Gaussian Processes}
\label{subsec:lgp}

\begin{figure}[htbp]
    \centering
    \begin{tikzpicture}[
    node distance=1.5cm,
    every node/.style={transform shape},
    gp1/.style={circle, fill=green!20, minimum size=1cm},
    gp2/.style={circle, fill=red!20, minimum size=1cm},
    arr/.style={->, >=stealth, thick}
]

\node (x1) at (-2,2) {$\mathbf{X}$};
\node (x2) at (-2,0) {$\mathbf{X}$};
\node (x3) at (-2,-2) {$\mathbf{X}$};

\node[gp1] (gp11) at (0,2) {$\mathcal{GP}_1^{(1)}$};
\node[gp1] (gp12) at (0,0) {$\mathcal{GP}_1^{(2)}$};
\node[gp1] (gp13) at (0,-2) {$\mathcal{GP}_1^{(P)}$};

\node[gp2] (gp2) at (3,0) {$\mathcal{GP}_2$};

\node (y) at (5,0) {Y};

\node at (0,-1) {$\vdots$};
\node at (-2,-1) {$\vdots$};

\draw[arr] (x1) -- (gp11);
\draw[arr] (x2) -- (gp12);
\draw[arr] (x3) -- (gp13);

\draw[arr] (gp11) -- node[above, sloped] {$\mathbf{W}_{*1}$} (gp2);
\draw[arr] (gp12) -- node[above] {$\mathbf{W}_{*2}$} (gp2);
\draw[arr] (gp13) -- node[below, sloped] {$\mathbf{W}_{*P}$} (gp2);

\draw[arr] (gp2) -- (y);
\end{tikzpicture}
    \caption{A two-layered deep Gaussian process model}
    \label{fig:dgp}
\end{figure}
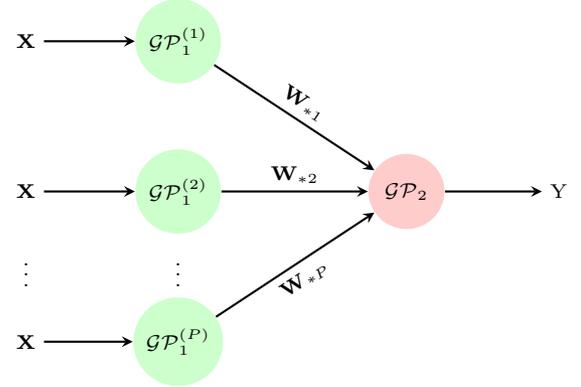

Consider a GP model with $N$ sets of $D$-dimensional input ($\mathbf{X} \in \mathbb{R}^{N \times D}$) and produces $N$ sets of $P$-dimensional output ($\mathbf{W} \in \mathbb{R}^{N \times P}$). We can assume that the output $\mathbf{W}$ of this model, i.e. the column vectors $\mathbf{W}_{*p}$, is conditionally independent with respect to $\mathbf{X}$. A linked GP (LGP) is then created when we link the output $\mathbf{W}$ to a second GP model that produces $N$ one-dimensional outputs ($\mathbf{Y} \in \mathbb{R}^{N}$). Let the GP surrogates of the two models be $\mathcal{GP}_1$ and $\mathcal{GP}_2$, respectively. Then, $\mathcal{GP}_1$ is a collection of independent GPs, $\{\mathcal{GP}_1^{(p)}\}_{p=1,...,P}$, and the hierarchy of GPs that represents the system can be seen as in Fig.~\ref{fig:dgp}. As described in Subsection~\ref{subsec:gp}, we can see that each GP corresponds to a multivariate Gaussian distribution with input $\mathbf{X}$ and output $\mathbf{W}_{*p}$.

Assume that the model parameters involved in $\mathcal{GP}_1$ and $\mathcal{GP}_2$ are known or estimated and we observe realisations $\mathbf{w}$ and $\mathbf{y}$ of $\mathbf{W}$ and $\mathbf{Y}$ given inputs $\mathbf{X} = \mathbf{x}$. Then, the posterior predictive distribution of the global output $Y_0$ at a new global input position $\mathbf{x}_0$ is given by $Y_0|\mathcal{D} \sim p(y_0|\mathbf{x}_0;\mathbf{y},\mathbf{w},\mathbf{x})$, where $\mathcal{D} = \{\mathbf{Y} = \mathbf{y}, \mathbf{W} = \mathbf{w}, \mathbf{X} = \mathbf{x}\}$ and $p(y_0|\mathbf{y},\mathbf{w},\mathbf{x})$ is the PDF of $Y_0|\mathcal{D}$. Note that
\begin{equation}
    \label{eq:pred}
    \begin{split}
    p(y_0|\mathbf{x}_0; \mathbf{y}, \mathbf{w}, \mathbf{x})
    &= \int p(y_0|\mathbf{w}_0;\mathbf{y},\mathbf{w},\mathbf{x})p(\mathbf{w}_0|\mathbf{x}_0;\mathbf{y},\mathbf{w},\mathbf{x})d\mathbf{w}_0 \\
    &= \int p(y_0|\mathbf{w}_0;\mathbf{y},\mathbf{w}) \prod_{p=1}^P p(w_{0p}|\mathbf{x}_0; \mathbf{w}_p^*,\mathbf{x}) d\mathbf{w}_0
    \end{split}
\end{equation}
where $p(y_0|\mathbf{w}_0;\mathbf{y},\mathbf{w})$ and $p(w_{0p}|\mathbf{x}_0;\mathbf{w}_p^*,\mathbf{x})$ are PDF's of the posterior predictive distributions of $GP_2$ and $GP_1^{(p)}$, respectively; and $\mathbf{w}_0 = (w_{01},\ldots,w_{0P})$. However, $p(y_0|\mathbf{x}_0;\mathbf{y},\mathbf{w},\mathbf{x})$ is analytically intractable because the integral in Equation~\ref{eq:pred} does not have a closed form expression. It can be shown that, given the GP specifications in Subsection~\ref{subsec:gp}, the mean, $\tilde{\mu}_0$, and variance, $\tilde{\sigma}^2_0$, of $Y_0|\mathcal{D}$ are expressed analytically as follows:
\begin{equation}
    \tilde{\mu}_0 = \mathbf{I}(\mathbf{x}_0)^T \mathbf{R}(\mathbf{w})^{-1}\mathbf{y}
\end{equation}
\begin{equation}
    \begin{split}
    \tilde{\sigma}^2_0 =~&\mathbf{y}^T \mathbf{R}(\mathbf{w})^{-1}\mathbf{J}(\mathbf{x}_0)\mathbf{R}(\mathbf{w})^{-1}\mathbf{y} - \left[\mathbf{I}(\mathbf{x}_0)^T \mathbf{R}(\mathbf{w})^{-1}\mathbf{y}\right]^2 \\
    &+ \sigma^2\left(1 + \eta - \text{tr}\left[\mathbf{R}(\mathbf{w})^{-1} \mathbf{J}(\mathbf{x}_0)\right]\right)
    \end{split}
\end{equation}
where $\mathbf{I}(\mathbf{x}_0) \in \mathbb{R}^{N \times 1}$ with its $i$th element
$I_i = \prod_{p=1}^P \mathbb{E}[k_p(W_{0p}(\mathbf{x}_0), w_{ip})]$; $\mathbf{J}(\mathbf{x}_0) \in \mathbb{R}^{N \times N}$ with its $ij$th element $J_{ij} = \prod_{p=1}^P \mathbb{E}[k_p(W_{0p}(\mathbf{x}_0), w_{ip}) k_p(W_{0p}(\mathbf{x}_0), w_{jp})]$; and the expectations in $\mathbf{I}(\mathbf{x}_0)$ and $\mathbf{J}(\mathbf{x}_0)$ have closed-form expressions under the linear kernel, squared exponential kernel, and a class of Matérn kernels \cite{titsias2010bayesian,kyzyurova2018coupling,ming2021linked}. The linked GP is then defined as a Gaussian approximation $\hat{p}(y_0|\mathbf{x}_0; \mathbf{y}, \mathbf{w}, \mathbf{x})$ with its mean and variance given by $\tilde{\mu}_0$ and $\tilde{\sigma}^2_0$. Furthermore, the linked GP can be built iteratively to analytically approximate the posterior predictive distribution of outputs from any feed-forward GP systems. Research has shown that this approach provides an adequate approximation by minimising the Kullback-Leibler divergence \cite{ming2021linked}.

\subsection{Deep Gaussian Processes}
\label{subsec:dgp}

A deep GP (DGP) model can be seen as a linked GP model when the internal inputs/outputs of GPs are latent. Thus, Fig.~\ref{fig:dgp} can be seen as a two-layered deep GP with unobserved values of variable $\mathbf{W}$. The latent variables make conducting efficient inference for deep GP models harder. For instance, to train the two-layer deep GP by the maximum likelihood approach, one needs to optimise the model parameters by maximising the likelihood function:
\begin{equation}
\label{eq:likelihood}
\mathcal{L} = p(\mathbf{y}|\mathbf{x}) = \int p(\mathbf{y}|\mathbf{w}) \prod_{p=1}^P p(\mathbf{w}_{*p}|\mathbf{x})d\mathbf{w},
\end{equation}
where $p(\mathbf{y}|\mathbf{w})$ is the multivariate Gaussian PDF of $\mathcal{GP}_2$ and $p(\mathbf{w}_{*p}|\mathbf{x})$ is the multivariate Gaussian PDF of $\mathcal{GP}_1^{(p)}$. However, due to the nonlinearity between $\mathbf{y}$ and $\mathbf{w}$, the integral with respect to the latent variable $\mathbf{w}$ in Equation~\ref{eq:likelihood} is analytically intractable. As we increase the depth of a deep GP, the number of such intractable integrals will also increase.

\subsection{Deep GP with Stochastic Imputation}
\label{subsec:dgpsi}

Recently, stochastic imputation (SI) was proposed to tackle the inference issue in deep GP~\cite{ming2023deep}, leveraging the fact that DGP and LGP are similar in structure. This approach provides a well-balanced trade-off between computational complexity and accuracy by combining the computational efficiency of variational inference \cite{salimbeni2017doubly} and the accuracy of a full Bayesian approach \cite{sauer2022active}. The key concept of SI is converting a DGP emulator to multiple LGP emulators, each representing a DGP realisation with imputed latent variables.

Given a similar DGP emulator hierarchy as described in Subsection~\ref{subsec:lgp} (Fig.~\ref{fig:dgp}) and realisations $\mathbf{y}$ of $\mathbf{Y}$, we can obtain point estimates of unknown model parameters in $\mathcal{GP}_1^{(p)}$ for all $p=1,\dots,P$ and $\mathcal{GP}_2$, using the Stochastic Expectation Maximization (SEM) algorithm~\cite{ming2023deep}. With the estimated model parameters, the DGP emulator gives the approximate
posterior predictive mean and variance of ${y}_0(\mathbf{x}_0)$ at a new input position $\mathbf{x}_0$ as described in Algorithm~\ref{alg:dgpsi}.

\begin{algorithm}[htbp]
    \caption{Construction of a DGP emulator with the hierarchy in Figure~\ref{fig:dgp}}
    \label{alg:dgpsi}
    \begin{algorithmic}[1]
    \Require{i) Realisations $\mathbf{x}$ and $\mathbf{y}$; ii) A new input position $\mathbf{x}_0$; iii) The number of imputations $N$.}
    \Ensure{Mean and variance of ${y}_0(\mathbf{x}_0)$.}
    \For{$i=1,\dots,N$}
    \State{\label{alg:one_ess}Given $\mathbf{x}$ and $\mathbf{y}$, draw an imputation $\{\mathbf{w}_{*p,i}\}_{p=1,\dots,P}$ of the latent output $\{\mathbf{W}_{*p}\}_{p=1,\dots,P}$ via an Elliptical Slice Sampling~\cite{nishihara2014parallel} update.}
    \State{Construct the LGP emulator $\mathcal{LGP}_i$ with the mean $\tilde{\mu}_{0,i}(\mathbf{x}_0)$ and variance ${\tilde{\sigma}_{0,i}^2}(\mathbf{x}_0)$, given $\mathbf{x}$, $\mathbf{y}$, and $\{\mathbf{w}_{*p,i}\}$.}
    \EndFor
    \State{Compute the mean ${\mu}(\mathbf{x}_0)$ and variance ${\sigma}^2(\mathbf{x}_0)$ of ${y}_0(\mathbf{x}_0)$ by
    \begin{align*}
     {\mu}(\mathbf{x}_0)&=\frac{1}{N}\sum_{i=1}^N\tilde{\mu}_{0,i}(\mathbf{x}_0),\\
     {\sigma}^2(\mathbf{x}_0)&=\frac{1}{N}\sum_{i=1}^N\left([\tilde{\mu}_{0,i}(\mathbf{x}_0)]^2+{\tilde{\sigma}_{0,i}^2}(\mathbf{x}_0)\right)-{\mu}(\mathbf{x}_0)^2.
    \end{align*}
    }
    \end{algorithmic} 
\end{algorithm}

One can extend Algorithm~\ref{alg:dgpsi} to multiple layers $l = 1, \dots, L$ and multiple outputs $\mathbf{y}_0(\mathbf{x}_0)$ by applying the same algorithm and repeating step 2 for each layer. A detailed explanation of this generalisation is provided in \cite{ming2023deep}.

\subsection{Benchmarking}

Our numerical experiments evaluated five different models to analyse the dataset.

\paragraph{Linked GP with sequential design} This hierarchical approach consisted of three steps. First, we fitted individual GPs for each covariate using timestamps and removing observations with missing values (i.e. complete case analysis). Second, we trained a GP to predict the output variable using these covariates. Finally, we connected these GPs into a hierarchical structure, allowing output variable prediction even when covariate data is missing.

\paragraph{Deep GP with stochastic imputation} In contrast to the separated training approach, we trained a unified deep GP model that integrated all components: it took timestamp as input, processed covariates in the latent layer, and predicted the output variable as output|all within a single end-to-end framework as described in Subsection~\ref{subsec:dgpsi}.

To set the baseline for our proposed models, we compared them with the following approaches:
\begin{enumerate}
    \item Last-known value imputation: we used the last-known value of the variable until the next measurement is observed~\cite{siddiqui1998locf};
    \item MICE: multiple imputation using chained equations \cite{jarrett2022hyperimpute}|ignoring the temporal dependency, relying only on the cross-sectional information between variables by treating each observation as independent and identically distributed~\cite{rubin1987mice,little2019statistical}, and doing iterative imputation;
    \item GP interpolation: we fitted a GP regressor with a squared exponential kernel individually for each covariate and the output variable.
\end{enumerate}
\section{Numerical Experiments}
\label{sec:results}

\subsection{Clinical Problem}
\label{subsec:clinical-problem}

Building on the Stewart-Fencl approach, the model was constructed with pH as the dependent variable and three independent variables: CO$_2$ levels, strong ion differences (SID), and weak acid concentrations. As such, two experiments were conducted in this study:
\begin{enumerate}
    \item Development of a continuous pH estimation model based on the Stewart-Fencl approach~\cite{stewart1983modern}.
    \item Simulation of real-world clinical scenarios by deliberately masking (intentionally withholding) portions of the covariates, mirroring the practical challenges of aligning laboratory test results with continuous bedside monitoring. Using measured pH levels to impute these missing covariate values, the aim was to provide clinicians with insights into patient status between laboratory tests.
\end{enumerate}

The experiments were done on a dataset of 14 ICU admission windows selected at random from a paediatric intensive care unit. Each admission had a different number of data points, ranging from 19 to 115 hourly timestamps, and some patients had multiple admissions. To further de-identify the patients, the dates were shifted to future dates while retaining the time relationships.

To follow the physicochemical approach of acid-base balance, the model incorporated multiple blood gas measurements. Specifically, the model used partial carbon dioxide pressure (pCO$_2$) and pH measurements and the difference between Na$^+$ and Cl$^-$ concentrations to represent the SID~\cite{kellum2000ph}. Although CO$_2$ levels can be measured through both blood gas analysis~\cite{hassan2024arterial} and capnography (ETCO$_2$)~\cite{raffe2020oximetry}, the numerical experiment was simplified by using only blood gas measurements.
For acid-base balance modelling, blood gas measurements are generally preferred over capnography for CO$_2$ because they offer a more comprehensive and direct assessment of the body's acid-base status~\cite{anderson2000co2,lee2024pitfalls}. Additionally, due to limited observations in half of the admission windows, the weak acid component was represented by lactate measurements from the blood gas analyser.

\begin{figure}[htbp]
    \centering
    \begin{tikzpicture}[
    node distance=1.5cm,
    every node/.style={transform shape},
    gp1/.style={circle, fill=green!20, minimum size=1cm},
    gp2/.style={circle, fill=red!20, minimum size=1cm},
    arr/.style={->, >=stealth, thick}
]

\node (x1) at (-2,0) {$\mathbf{time}$};

\node[gp1] (gp11) at (0,2) {$\mathcal{GP}_1^{(1)}$};
\node[gp1] (gp12) at (0,0) {$\mathcal{GP}_1^{(2)}$};
\node[gp1] (gp13) at (0,-2) {$\mathcal{GP}_1^{(3)}$};

\node[gp2] (gp2) at (3,0) {$\mathcal{GP}_2$};

\node (y) at (5,0) {\textbf{pH}};

\draw[arr] (x1) -- (gp11);
\draw[arr] (x1) -- (gp12);
\draw[arr] (x1) -- (gp13);

\draw[arr] (gp11) -- node[above, sloped] {\textbf{pCO$_2$}} (gp2);
\draw[arr] (gp12) -- node[above] {\textbf{SID}} (gp2);
\draw[arr] (gp13) -- node[below, sloped] {\textbf{Lactate}} (gp2);

\draw[arr] (gp2) -- (y);

\draw[gray] (-2.5,-3) rectangle (2.4,3);  
\draw[gray] (-1,-2.8) rectangle (5.5,2.8);  

\node at (-2.3,2.7) {1};  
\node at (-0.8,2.5) {2};  

\end{tikzpicture}
    \caption{A two-layered deep Gaussian process (DGP) for pH prediction. This study compares two fitting approaches: the linked GP with sequential design (LGP), which fits the layers separately (first predicting the covariates, then pH) and the DGP with stochastic imputation (SI) method, where all the components are trained simultaneously.}
    \label{fig:dgp-ph}
\end{figure}
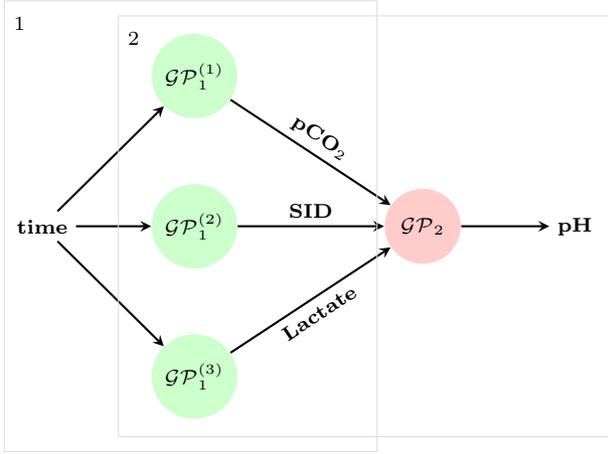

The DGP-SI model was trained simultaneously for all components using the architecture illustrated in Fig.~\ref{fig:dgp-ph}. For comparison, an ablation study was also conducted using an LGP model with a sequential design. This linked model operates in two steps: (1) it used separate GPs to predict three covariates (pCO$_2$, SID, and lactate) from time inputs; (2) it combined these predicted covariates to forecast pH. For a fair comparison, the MICE model only used time and pH data, excluding covariates since these would not be available during the inference process of our proposed models.

Following the literature, the data was preprocessed through several steps. First, the data was discretised into sequences of hourly intervals and aggregated the measurements by taking the arithmetic means~\cite{lipton16rnn,ghosheh2024ignite}. Then, to evaluate the model's robustness, either the observed pH or the covariates were randomly masked with varying proportions (10\%, 20\%, 30\%, and 40\%)~\cite{jafrasteh2023mgp, beaulieu2017missing}. Finally, z-score transformation was applied to both pH values and the covariates to standardise the data distribution.


To evaluate model performance given the presence of measurement noise, the accuracy of missing value imputations was defined as the mean absolute error (MAE):
\begin{equation}
    MAE = \frac{1}{N \times D} \sum_{i=1}^N \sum_{d \in D} |X_{id} - \hat{X}_{id}|
\end{equation}
where $N$ is the total number of missing values being evaluated, $X_{id}$ is the $i$-th true value that was masked and $\hat{X}_{id}$ is the $i$-th estimated value at dimension $d$. In the GP-based methods, the model prediction is computed as the mean of the predictive distribution.

\subsection{Results}
\label{subsec:results}

\begin{figure}[htbp]
    \centering
    \includegraphics[width=0.9\linewidth]{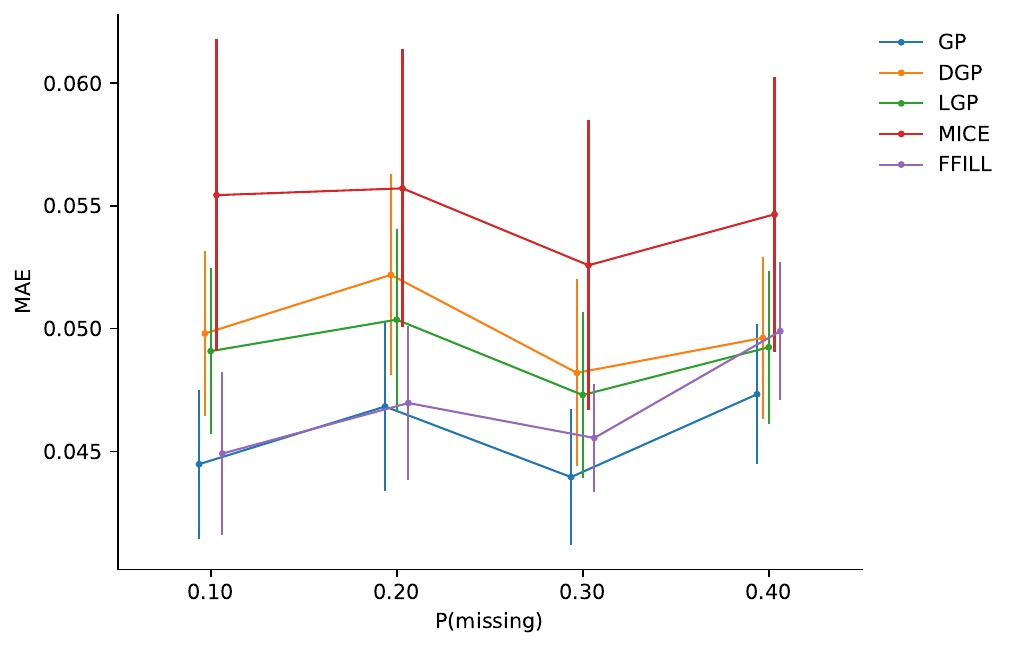}
    \caption{Average MAE values in predicting pH values. The error bar represents the standard error of the MAE values across all admission windows.}
    \label{fig:mae-ph}
\end{figure}

\paragraph{Predicting pH values}

To ensure comparable results, all models were standardised to use time as the input variable and pH as the target variable. While predicting pH values, the model only used time as the predictor variable and information from other covariates was excluded. Additionally, the last-known imputation method only used the most recent pH measurements, while in the single GP fitting, pH measurements before and after the missing data points were interpolated. Although the DGP and LGP models had access to covariate information as the latent variables during their development, they also operated with time as their only input variable during the inference process.

The analysis of prediction errors revealed a hierarchy among the imputation methods of pH values. As shown in Fig.~\ref{fig:mae-ph}, the GP interpolation outperformed other methods, maintaining the lowest error rates even as the proportion of missing values increased. Following closely, the last-known imputation method had comparable average errors to the GP interpolation when the proportion of missingness is 10\% or 20\%, but showed declining accuracy at higher proportions. DGP and LGP models ranked in the middle, performing better than MICE but worse than both GP interpolation and last-known imputation. Their reduced accuracy might stem from uncertainty propagation through their latent variables. MICE, which ignores temporal relationships by treating timesteps independently, consistently showed the highest error rates.

\begin{figure}[htbp]
    \centering
    \includegraphics[width=0.9\linewidth]{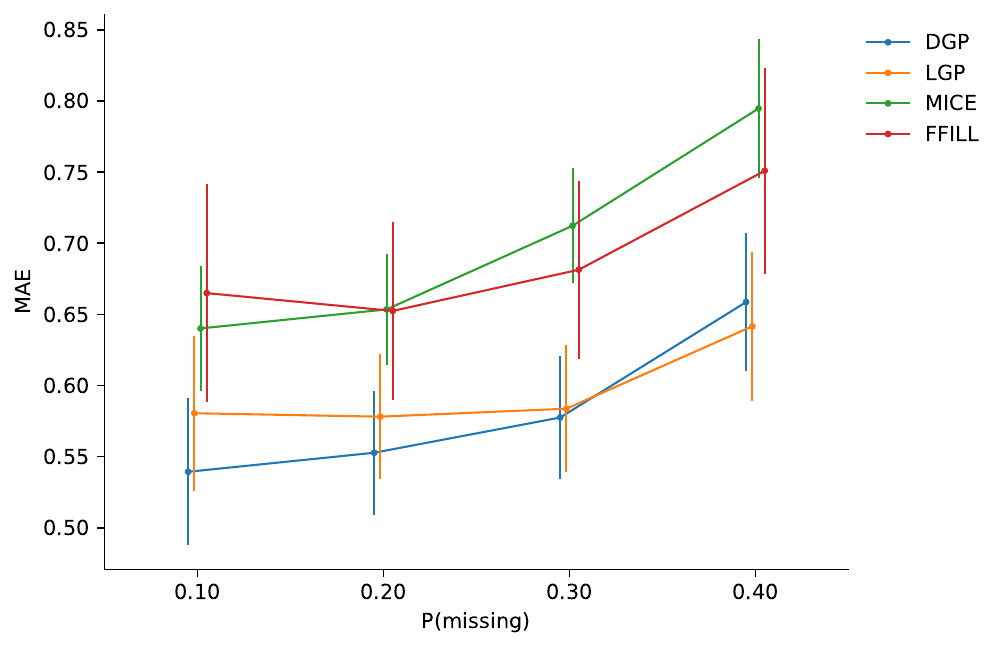}
    \caption{Average MAE values in imputing the covariates. The error bar represents the standard error of the MAE values across all admission windows. DGP, which combines longitudinal and cross-sectional information, performs better in imputing missing values in the covariates, particularly at lower missingness levels. However, as the proportion of missing values increases, methods that rely on longitudinal information become more effective.}
    \label{fig:mae-features}
\end{figure}

\paragraph{Imputing missing values in covariates}

Unlike the pH value prediction, where missing values were filled out from a single variable, the second experiment predicted missing values with varying proportions from three variables: pCO$_2$, SID, and lactate as the weak acid. All three variables were measured at the same time as pH. This experiment used all the observed pH values to link the three covariates as suggested by the Stewart-Fencl physicochemical approach. Since the linked GP, in this case, was just individually fitted GPs with complete case analysis, there were only four models to compare in this experiment.

The performance comparison revealed that DGP achieved the lowest error rates at 10\% and 20\% missing values, performed similarly to LGP at 30\%, and slightly underperformed compared to LGP at 40\% (Figure~\ref{fig:mae-features}). Both DGP and LGP demonstrated better performance than MICE and last-known imputation methods, with last-known imputation showing lower error rates than MICE as missingness increased. These findings suggest that longitudinal information is more valuable than cross-sectional information for covariate imputation. However, DGP's low error rates indicate that combining both longitudinal and cross-sectional information yields optimal results.

\paragraph{Uncertainty quantification}

\begin{figure}
    \centering
    \includegraphics[width=0.5\linewidth]{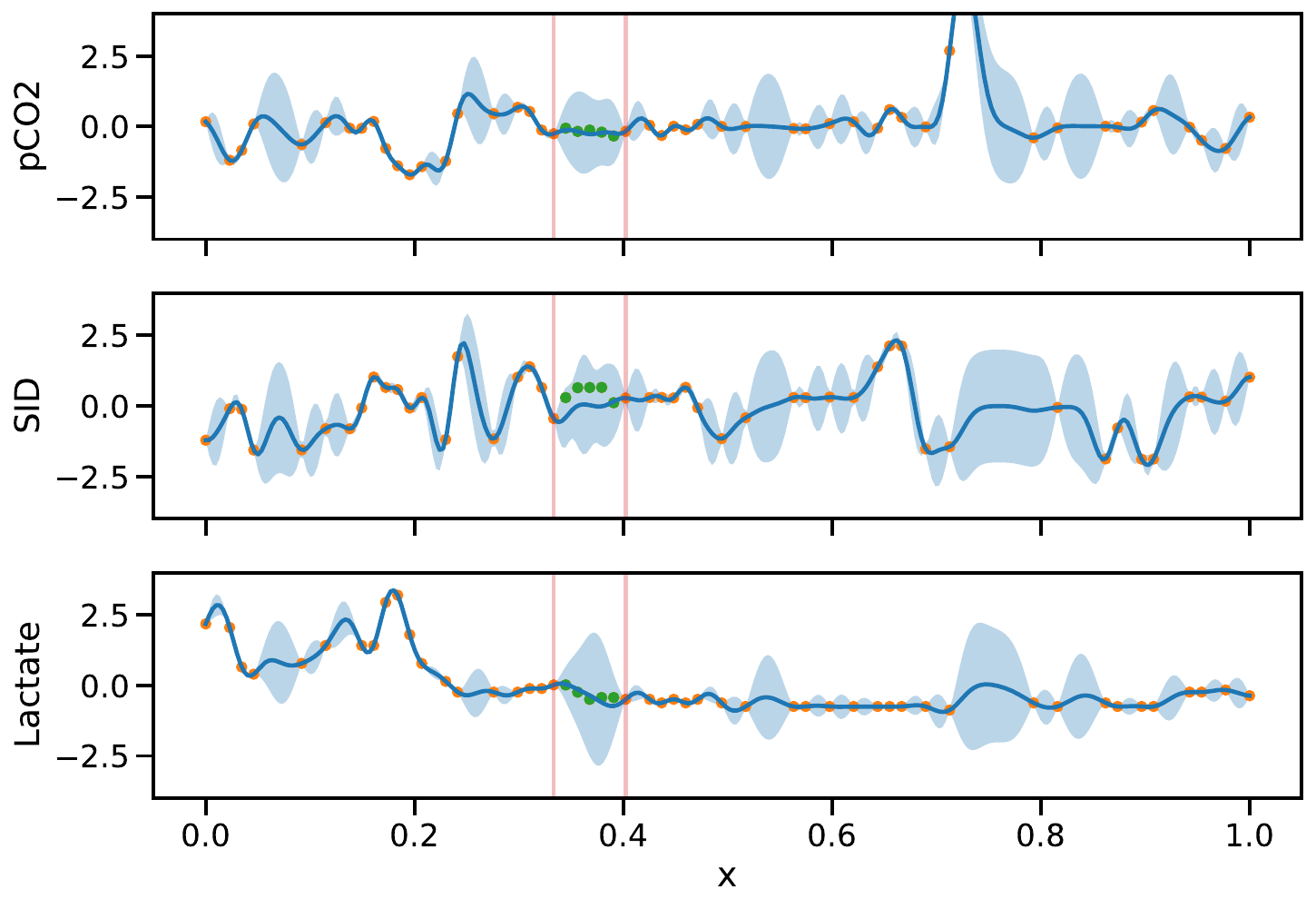}%
    \includegraphics[width=0.5\linewidth]{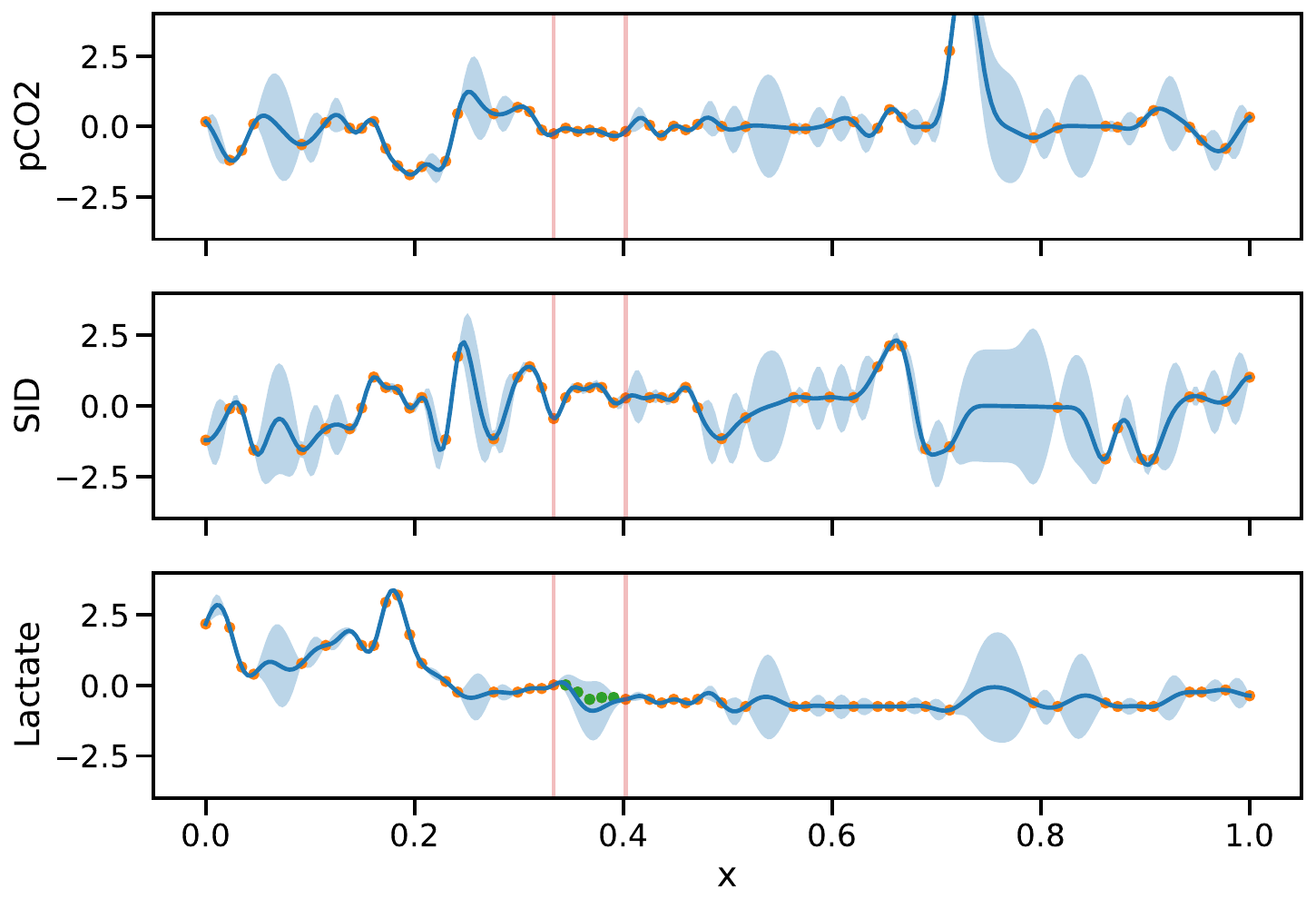}
    \includegraphics[width=0.5\linewidth]{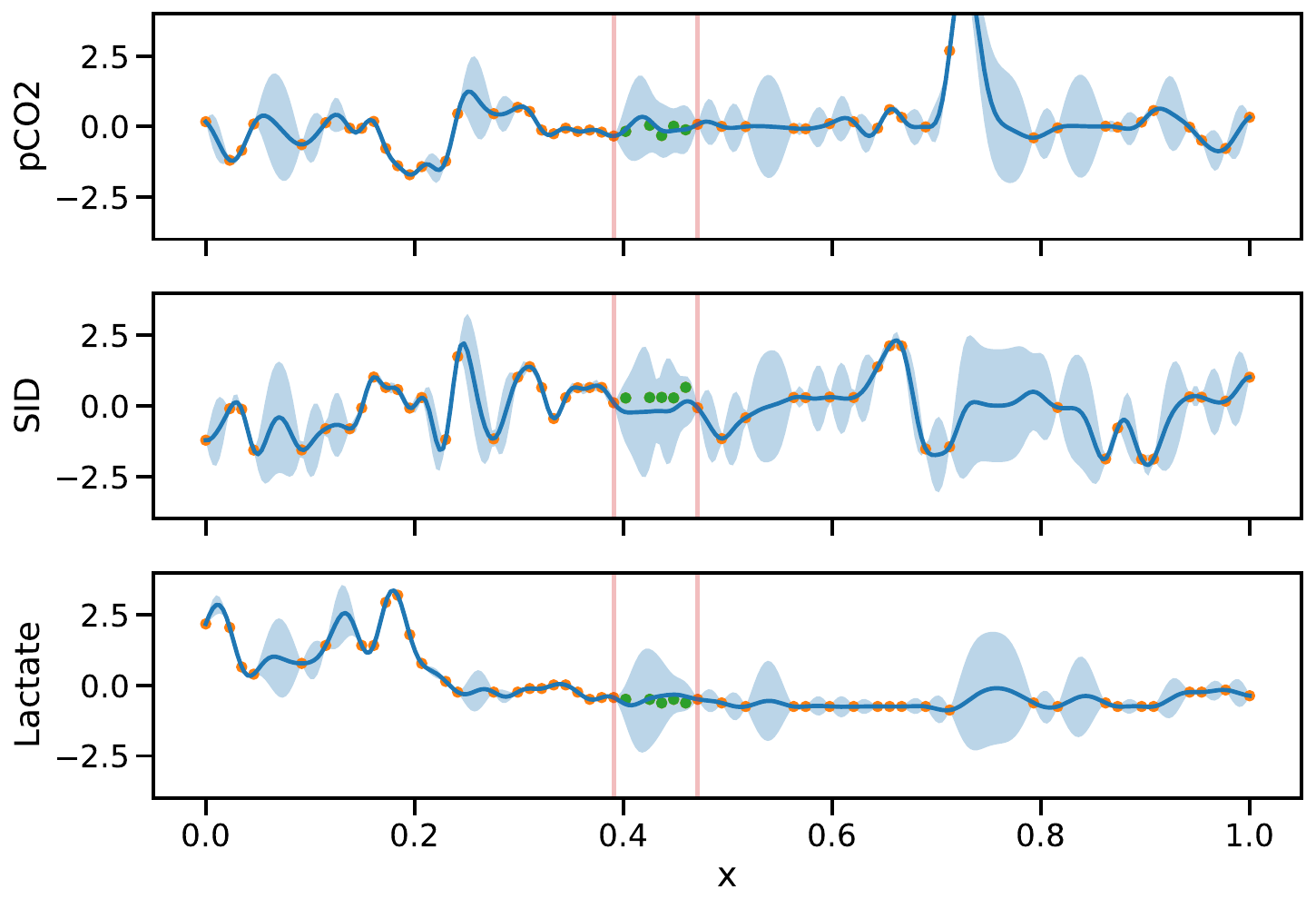}%
    \includegraphics[width=0.5\linewidth]{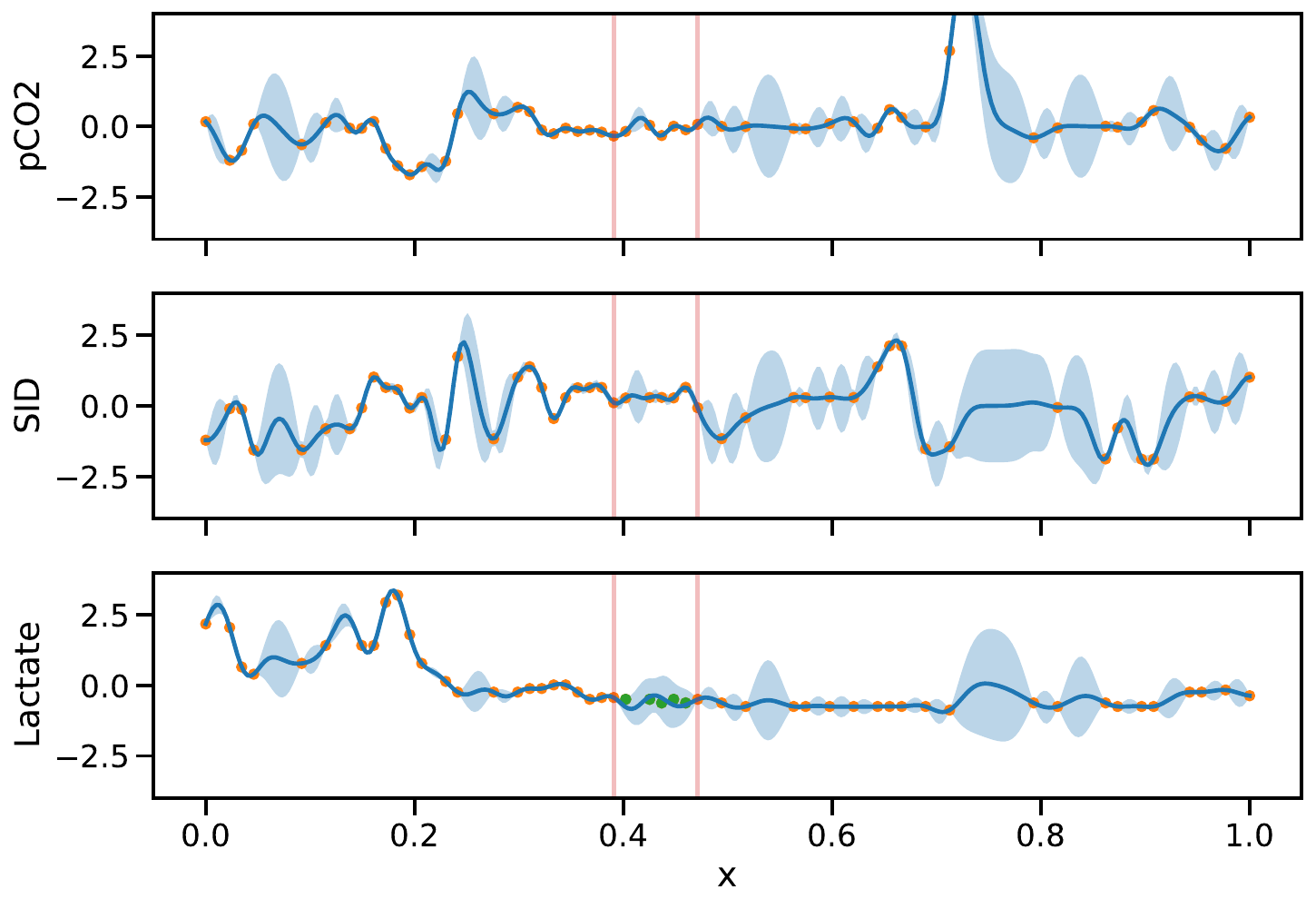}
    \includegraphics[width=0.5\linewidth]{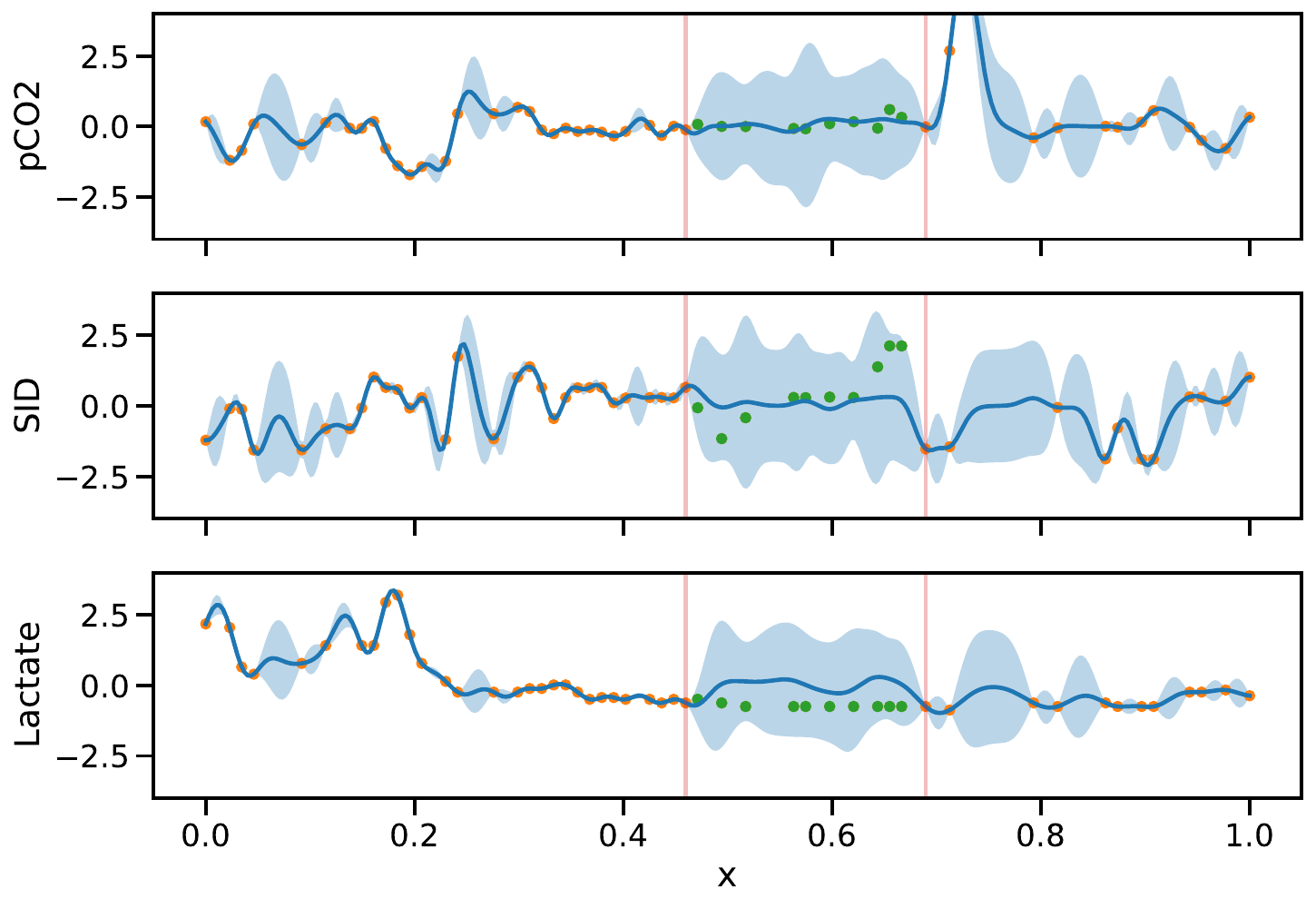}%
    \includegraphics[width=0.5\linewidth]{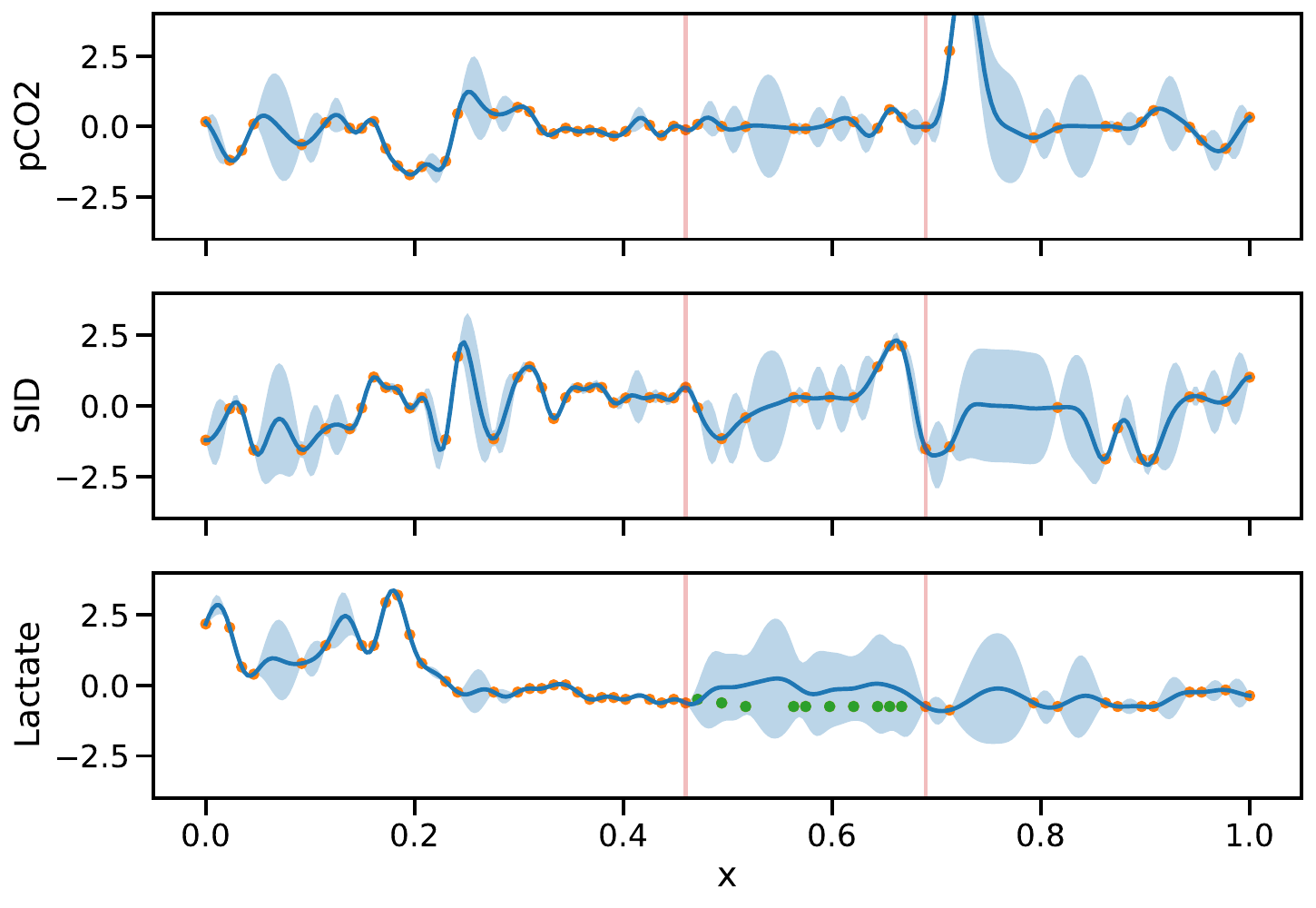}
    \caption{Uncertainty quantification derived from the DGP-SI model when imputing missing values in the covariates. Orange dots represent observed values, while green dots (between red vertical lines) show masked values. The x-axis shows scaled time, and the y-axis shows standardised covariate values. The uncertainty for lactate resulting from manually masking the three covariates (left) is greater than that from only masking lactate (right).}
    \label{fig:uncertainty}
\end{figure}

As the covariates were connected through pH in the output layer using DGP-SI, an observation from one covariate could affect the uncertainty of another covariate where an observation was unavailable. To demonstrate this effect, differences in uncertainty were compared by manually masking observations from the three covariates in three different intervals, focusing on masking lactate within these intervals. The experiment revealed that the uncertainty in lactate, shown in Fig.~\ref{fig:uncertainty}, was less when only the observed points in lactate were masked instead of all three covariates being masked.

\section{Discussion}
\label{sec:conclusion}

This paper demonstrated that DGP-SI, initially developed for uncertainty quantification in computationally expensive models, could effectively handle missing values in critical care data from different sources. The analysis across 14 admission windows showed that DGP-SI performed better in imputing missing covariate values, particularly when the proportion of missing data was low. This approach offers clinicians insight into patient states between measurements whilst providing uncertainty quantification, hence attaching a measure of confidence that addresses the inherent uncertainties in medical science~\cite{cabitza2017unintended}.

From a clinical perspective, accurately imputing missing values with quantified uncertainty can impact decision-making, especially in time-sensitive critical care scenarios. Clinicians often face the challenge of making treatment decisions with incomplete data, and the uncertainty quantification provided by DGP-SI could be valuable for retrospective analyses of acid-base disorders where multiple parameters interact in complex ways that clinicians find difficult to intuit. Additionally, this method has the potential to enhance early warning systems in intensive care units by providing more complete data streams for continuous patient monitoring, identifying deteriorating patients earlier while reducing false alarms.

The proposed approach also poses challenges for future work with two main limitations. First, it is less effective in emulating the Stewart-Fencl physicochemical model for pH prediction, likely due to error propagation through intermediate variables. Second, as highlighted in~\cite{ming2023deep}, the method becomes computationally expensive with larger datasets. To handle this, the data was partitioned into shorter admission windows. However, such an approach may not be feasible in settings with high-frequency measurements, such as those from wearable devices.

To address these limitations, several paths forward exist. The computational burden can be reduced through time discretisation and data aggregation, as demonstrated in this work. Alternative solutions include implementing sparse GP approximations~\cite{snelson2007local,bauer2016understanding} or utilising GPU parallelisation for exact GP computations~\cite{wang2019exact}. Additionally, further research should evaluate the model's performance on multimodal datasets with naturally varying patterns of missing data, focusing on scenarios where sufficient observations exist to be emulated using a deep GP model.


\section{Competing interests}
No competing interest is declared.

\section{Author contributions statement}

A.A.S., D.M., and F.A.D. conceived the experiment(s), A.A.S. conducted the experiment(s), A.A.S., D.M., and F.A.D. analysed the results. A.A.S. wrote the original manuscript. D.M., F.A.D., T.J., and S.R. reviewed and edited the manuscript.

\section{Acknowledgements}

This work was supported by the UK Engineering and Physical Sciences Research Council [EP/W523835/1 to A.A.S.].

\bibliographystyle{plain}
\bibliography{main}

\end{document}